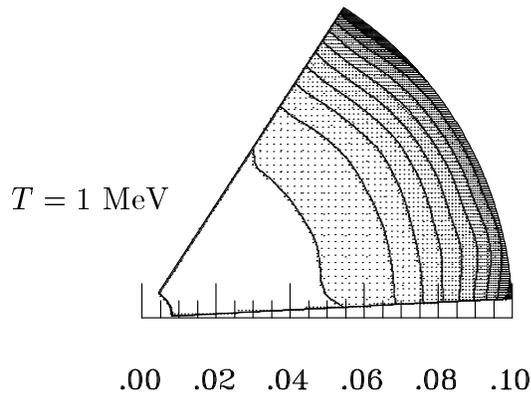
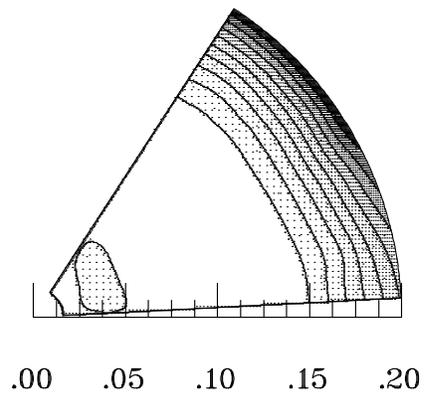

$T = 1$ MeV

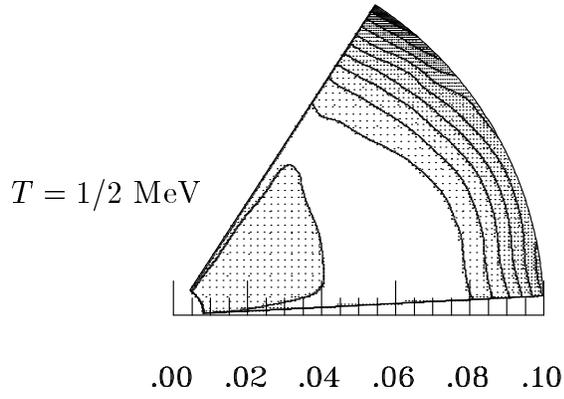
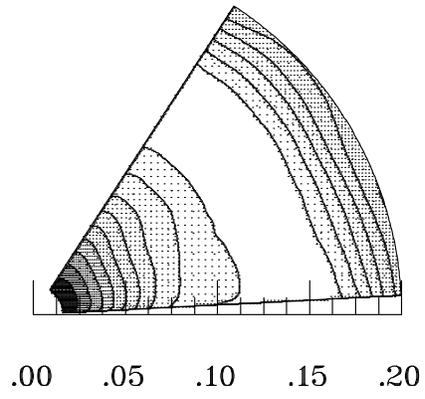

$T = 1/2$ MeV

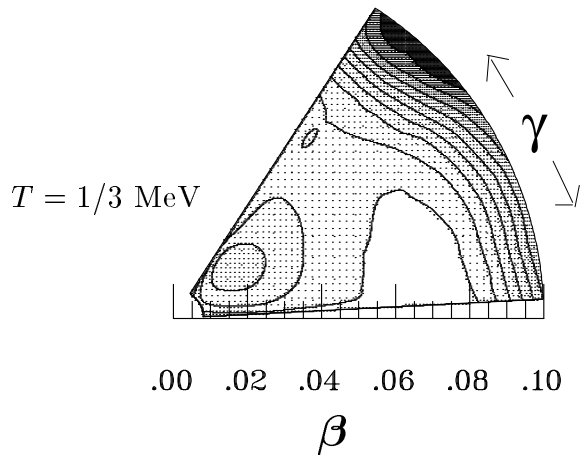
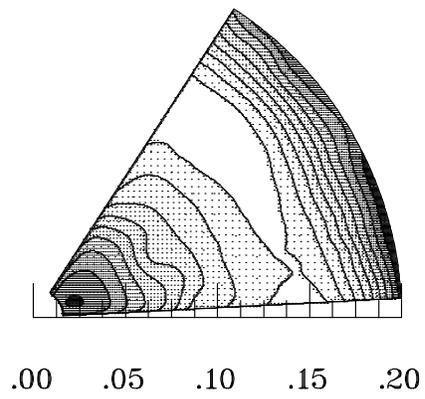

$T = 1/3$ MeV

$\gamma$

$\beta$

$^{128}$Te $\qquad\qquad\qquad ^{124}$Xe

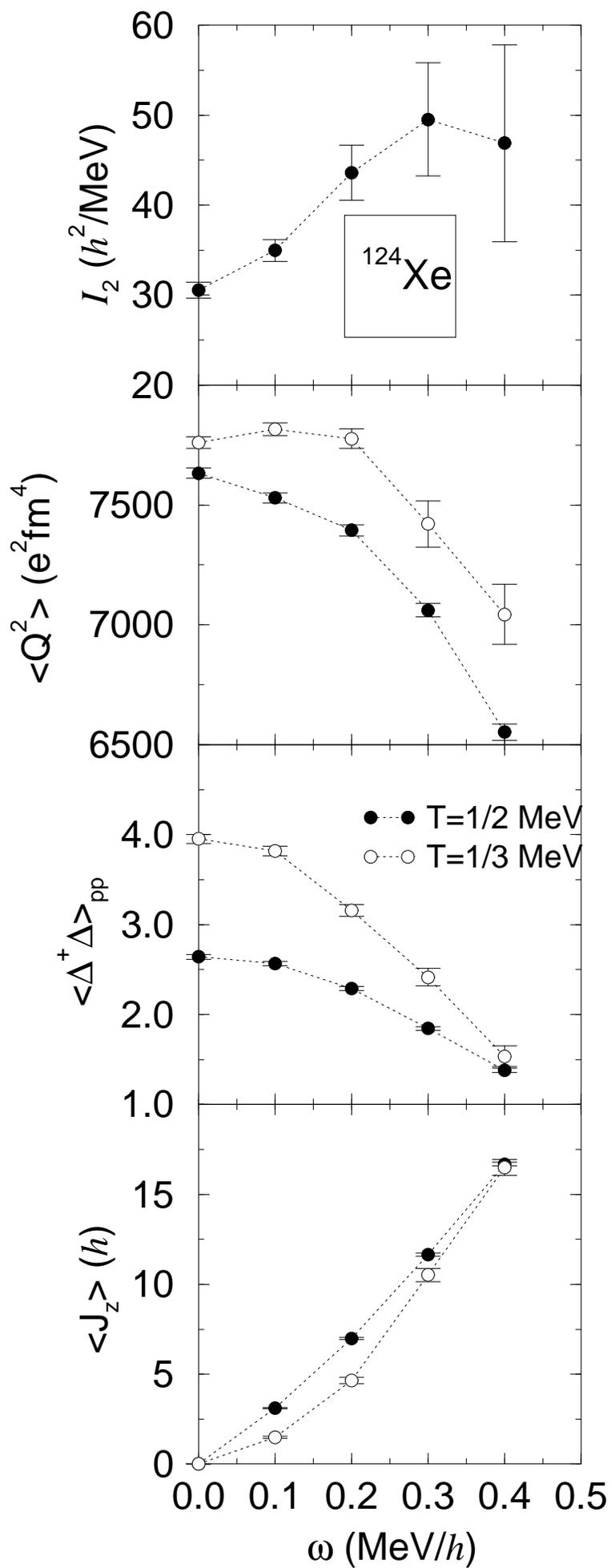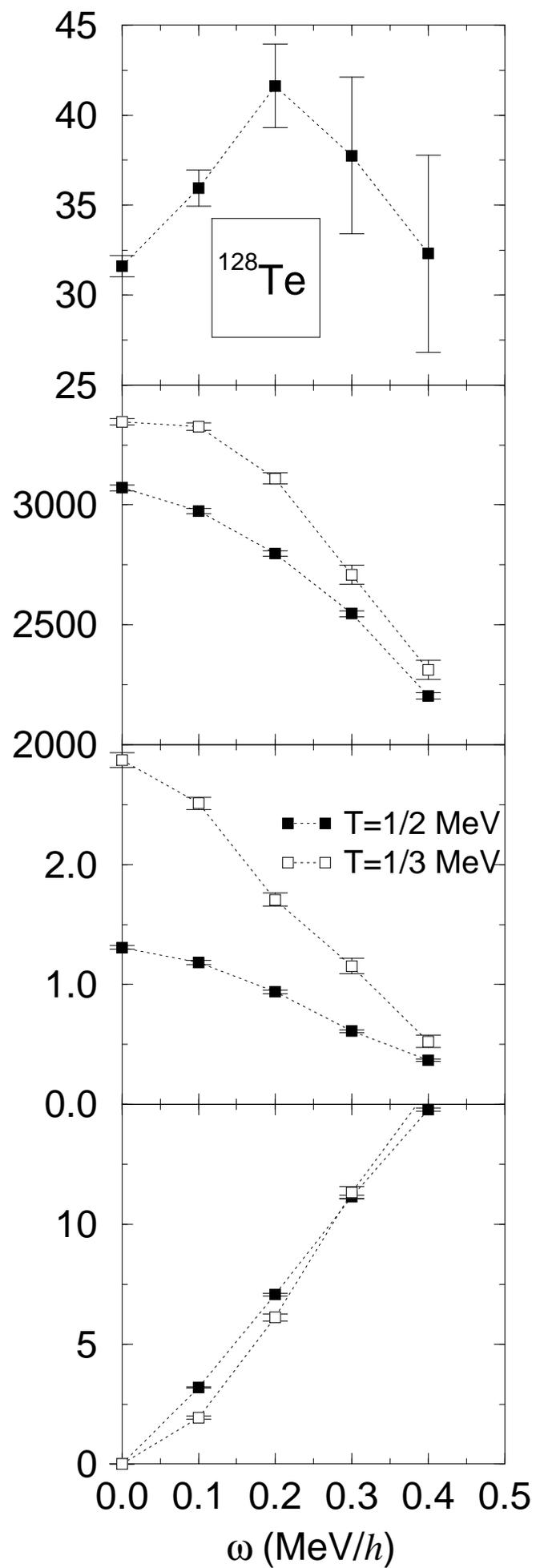

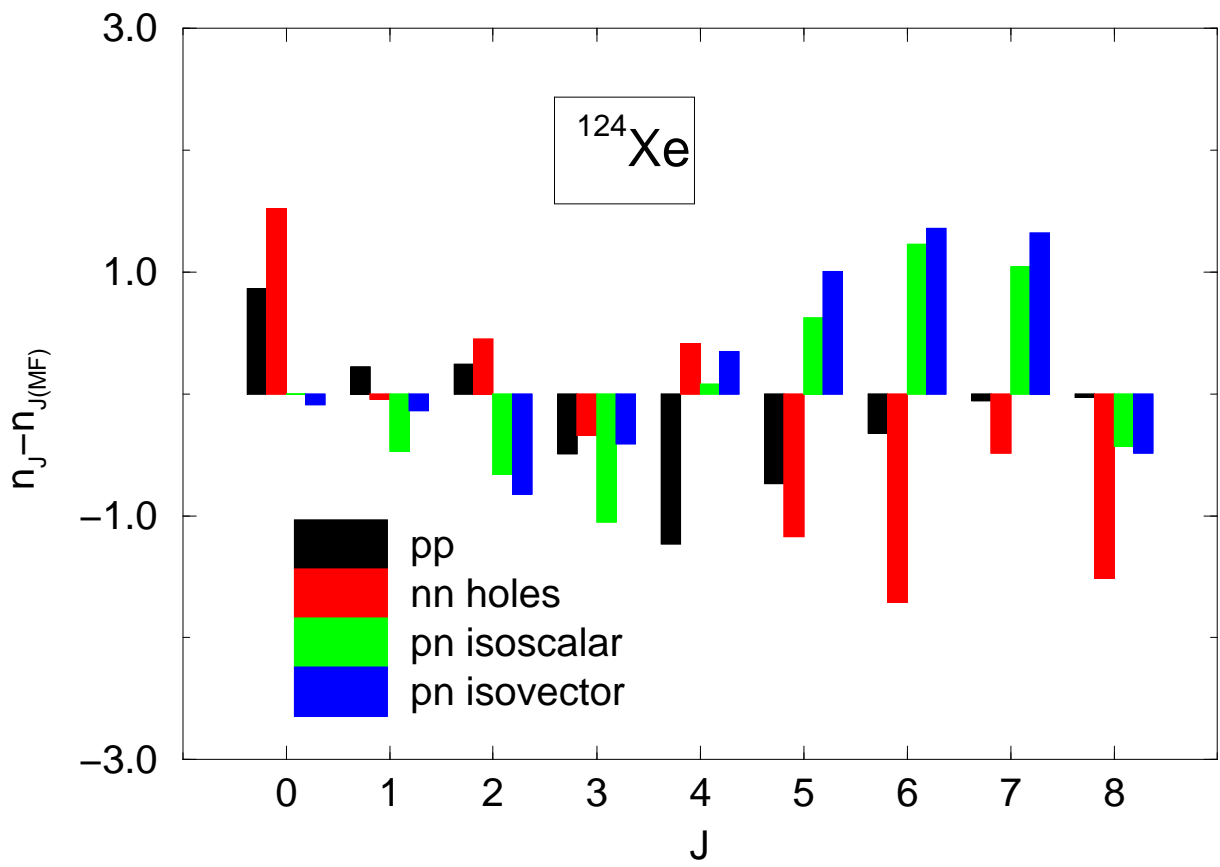

# Shell Model Monte Carlo Studies of $\gamma$-Soft Nuclei

Y. Alhassid[1], G.F. Bertsch[2], D.J. Dean[3] and S.E. Koonin[3]
[1] *Center for Theoretical Physics, Sloane Physics Laboratory,*
*Yale University, New Haven, CT 06520, USA*
[2] *Department of Physics, FM-15, University of Washington, Seattle, WA 98195 USA*
[3] *W. K. Kellogg Radiation Laboratory, 106-38, California Institute of Technology*
*Pasadena, California 91125 USA*
(July 31, 1995)

We present Shell Model Monte Carlo calculations for nuclei within the full major shell 50-82 for both protons and neutrons. The interaction is determined solely by self-consistency and odd-even mass differences. The methods are illustrated for $^{124}$Sn, $^{128}$Te and $^{124}$Xe. We calculate shape distributions, moments of inertia and pairing correlations as functions of temperature and angular velocity. Our calculations are the first microscopic evidence of $\gamma$-softness of nuclei in this region.

Nuclei with mass number $100 \leq A \leq 140$ are believed to have large shape fluctuations in their ground states. Associated with this softness are spectra with an approximate $O(5)$ symmetry and bands with energy spacings intermediate between rotational and vibrational. In the geometrical model these nuclei are described by potential energy surfaces with a minimum at $\beta \neq 0$ but independent of $\gamma$ [1]. Some of these nuclei have been described in terms of a quartic five-dimensional oscillator [2]. In the Interacting Boson Model (IBM) they are described by an $O(6)$ dynamical symmetry [3–5].

Nuclei with $100 \leq A \leq 140$ fill the major shell between 50 and 82 for both protons and neutrons, and conventional shell model calculations for the nuclei of interest in the full space are impossible. However, with the introduction of Shell Model Monte Carlo techniques (SMMC) [6], it has become possible to do exact calculations (up to a statistical error) in much larger model spaces [7] at zero and finite temperatures. This Letter presents the first fully microscopic calculations for soft nuclei with $100 \leq A \leq 140$ and compares them with the results of more phenomenological models.

An important problem is the choice of the interaction. We used a pairing plus quadrupole interaction, where the pairing contains both monopole and quadrupole terms whose strengths are determined by odd-even mass differences. The quadrupole interaction is derived from a surface-peaked separable force [8] with a strength fixed by self-consistency. Such an interaction is expected to be a reasonable way of describing deformation and pairing phenomena. A major advantage of the interaction we construct is that it has a "good" Monte Carlo sign, so that calculations can be done quite accurately without resorting to extrapolation techniques [9].

For the two-body interaction we take a multipole pairing force [10] plus an (isoscalar) self-consistent surface-peaked interaction. The latter is assumed to be separable $v(\mathbf{r}, \mathbf{r}') = -\chi (dV/dr)(dV/dr')\delta(\hat{r} - \hat{r}')$, where $V(r)$ is the (spherical) mean-field potential. The angular delta function is expanded in multipoles and only its quadrupole component is kept. We have

$$H_2 = -\sum_{\lambda\mu} \frac{\pi g_\lambda}{2\lambda + 1} P^\dagger_{\lambda\mu} P_{\lambda\mu} - \frac{1}{2}\chi : \sum_\mu (-)^\mu Q_\mu Q_{-\mu} :, \quad (1)$$

where :: denotes normal ordering and $P^\dagger_{\lambda\mu}$, $Q_\mu$ are pair and quadrupole operators given by

$$P^\dagger_{\lambda\mu} = \sum_{ab} (-)^{\ell_b} (j_a \parallel \mathcal{Y} \parallel j_b)[a^\dagger_{j_a} \times a^\dagger_{j_b}]_{\lambda\mu}$$
$$Q_\mu = -\frac{1}{\sqrt{5}} \sum_{ac} (j_a \parallel \frac{dV}{dr}\mathcal{Y}_{2\mu} \parallel j_c)[a^\dagger_{j_a} \times \tilde{a}_{j_c}]_{2\mu} . \quad (2)$$

In (2) $a \equiv n\ell j$ denotes a single particle orbit and $\tilde{a}_{jm} = (-)^{j+m} a_{j-m}$. The quadrupole interaction strength is determined self-consistently. A change in the mean-field potential is related to a change in the one-body density $\rho(\mathbf{r})$ through $\delta V(\mathbf{r}) = \int d\mathbf{r}' v(\mathbf{r}, \mathbf{r}')\delta\rho(\mathbf{r}')$. Using the invariance of the one-body potential under a displacement of the nucleus, and the separable form of the two-body interaction, we obtain

$$\chi^{-1} = \int_0^\infty dr\, r^2 \frac{dV}{dr}\frac{d\rho}{dr} . \quad (3)$$

The spherical nuclear density in (3) is calculated from $\rho(r) = (4\pi)^{-1} \sum_a f_a R_a^2(r)/r^2$, where $f_a$ and $R_a$ are the occupation number and the radial wavefunction of orbit $a$, respectively, and the sum goes over both the core and the valence shells. The self-consistent result (3) gives $\chi = 0.053$ MeV$^{-1}$ fm$^2$ for $^{124}$Xe. Since the interaction (1) is taken only in the valence shell, the value of $\chi$ found in (3) has to be renormalized. For a deformed harmonic oscillator single-particle potential, the core and valence quadrupole moments are equal [12], and the renormalization factor is 2. However, we use a Woods-Saxon potential $V(r)$, and find that a renormalization factor of $\sim 3$ is required to reproduce correctly the variation of the excitation energy of the first $2^+$ state in this regime.

The single-particle energies are determined from a Woods-Saxon plus spin-orbit potential

$$h_1 = -\frac{\hbar^2}{2m}\Delta + V(r) + \lambda_{ls}(\vec{\ell}\cdot\vec{s})r_0^2\frac{1}{r}\frac{dV}{dr} + \frac{1-\tau_z}{2}V_C(r)\,, \quad (4)$$

where $V(r) = V_0\left[1+\exp(r-R_0)/a\right]^{-1}$ with $R_0 = r_0 A^{1/3}$, and a Coulomb potential $V_C$ of a uniformly charged sphere is included for a proton ($\tau_z = -1$). We choose the parametrization of Ref. [11], where $V_0 = -49.6\left(1\pm 0.86\frac{N-Z}{A}\right)$ MeV for protons and neutrons, respectively; $a = 0.7$ fm; $r_0 = 1.275$ fm for protons and $r_0 = 1.347$ fm for neutrons. The spin-orbit potential has $\lambda_{\ell s} = 0.488$, $r_0 = 1.32$ fm and the same diffuseness parameter as the central part. For $^{124}$Xe, the resulting single-particle energies of $0g_{7/2}, 1d_{5/2}, 0h_{11/2}, 2s_{1/2}, 1d_{3/2}$ are -3.05, -3.35, -1.07, -1.04, -0.615 MeV for protons and -11.79, -12.08, -9.53, -10.21, -9.94 MeV for neutrons. Notice that the Coulomb potential has the effect of placing the proton's $h_{11/2}$ orbit below the $d_{3/2}$ and $s_{1/2}$ orbits. When the central Woods-Saxon potential is used for $V(r)$ in (2), we find that the corresponding matrix elements of the quadrupole interaction in the proton single-particle basis differ by only a few percent from those in the neutron single-particle basis. We can therefore choose either of these sets.

For the pairing interaction we include only monopole ($\lambda = 0$) and quadrupole ($\lambda = 2$) terms with $g_0 = g_2$. $g_0$ is determined in two steps. First, the pairing gap $\Delta$ is extracted from the experimental masses (or binding energies $\mathcal{B}$) of neighboring nuclei using [13] $\Delta_n = \frac{1}{4}[\mathcal{B}(N-2,Z)-3\mathcal{B}(N-1,Z)+3\mathcal{B}(N,Z)-\mathcal{B}(N+1,Z)]$. Next, using a particle-projected BCS approximation for the Hamiltonian (1), we find the value of $g_0$ that will reproduce the experimental gap for a spherical nucleus with the same mass number $A$. For $^{124}$Xe we find $g_0 = 0.150$ MeV. The inclusion of quadrupole pairing is important in order to bring down the excitation energy of the $2_1^+$ state in the tin isotopes to about 1.3 MeV (which is $2\Delta \sim 2$ MeV when only monopole pairing is included). Since both the monopole pairing and the quadrupole-quadrupole interaction are attractive, they satisfy the sign rule in the density decomposition and have a "good" Monte Carlo sign. The quadrupole pairing has components that violate that sign rule but they are all very small in comparison with the good sign components. Setting all the "bad" components to zero has no more than a 5% effect on the spectrum. In identifying the "bad" components of the interaction, one should take into account that orbits of both parities are present in the 50-82 shell, and the modified sign rule should be used [9]. Thus, we are able to derive a quasi-realistic interaction that has a "good" Monte Carlo sign.

We begin discussion of our results with the probability distribution of the quadrupole moment. This is an observable in principle, because it can be deduced from the expectation values of powers of the quadrupole operator. Observables in the SMMC are calculated by a weighted integration over their values for non-interacting nucleons moving in a fluctuating auxiliary field. We denote the single-particle evolution operator associated with an auxiliary field $\sigma$ by $U_\sigma$, and the quadrupole moment of each sample, $\langle Q_\mu\rangle_\sigma = Tr(Q_\mu U_\sigma)/Tr(U_\sigma)$, where $Q_\mu \equiv \sum r^2\mathcal{Y}_{2\mu}$. The method used in a recent SMMC study of deformed nuclei [7] follows the prescription $P(q_\mu) \propto \sum_\sigma \delta(\langle Q_\mu\rangle_\sigma - q_\mu)$. This method is valid provided $Tr(Q_\mu^n U_\sigma)/Tr(U_\sigma) = \langle Q_\mu\rangle_\sigma^n$. We can improve upon that prescription by explicitly forcing the quadratic moment to be correct. We calculate the variance of the Q operator for each sample, $\Delta_\sigma^2 = Tr(Q^2 U_\sigma)/Tr(U_\sigma) - \langle Q\rangle_\sigma^2$ and assume that higher cumulants vanish. Then

$$P(q_\mu) \propto \sum_\sigma \frac{1}{5\Delta_\sigma}\exp\left[-\frac{(\langle Q_\mu\rangle_\sigma - q_\mu)^2}{2\Delta_\sigma^2}\right]. \quad (5)$$

In order for (5) to define a shape distribution we must associate a deformation $\alpha_{2\mu}$ with each quadrupole moment $\langle Q_\mu\rangle_\sigma$. Starting from a deformed nucleus, $R = R_0(1+\sum_\mu \alpha_\mu \mathcal{Y}_{2\mu}^*)$, we calculate its quadrupole moment by expanding its density $\rho(\mathbf{r}) \equiv \rho_0(r-R)$ (where $\rho_0$ is the spherical density) to first order in deformation. We then find

$$\langle Q_\mu\rangle_\sigma = 4R_0\left(\int_0^\infty r^3 \rho_0(r)dr\right)\alpha_{2\mu}. \quad (6)$$

$\rho_0$ is calculated as below Eq. (3) but only within the valence shell. Eq. (6) is used to calculate $\alpha_{2\mu}$ from the measured $\langle Q_\mu\rangle_\sigma$. We then transform $\alpha_{2\mu}$ to the intrinsic frame where $\alpha_{20} = \beta\cos\gamma$ and $\alpha_{22} = \alpha_{2-2} = \beta\sin\gamma/\sqrt{2}$. The shape distribution $P(\beta,\gamma)$ can be converted to a free energy surface through $F(\beta,\gamma;T) = -T\log\left[P(\beta,\gamma)/\beta^3|\sin 3\gamma|\right]$, where the unitary metric $\prod_\mu \alpha_{2\mu} = \beta^4|\sin 3\gamma|\,d\beta d\gamma d\Omega$ has been assumed [14].

The shape distributions of $^{128}$Te and $^{124}$Xe are shown in Fig. 1 at different temperatures. These nuclei are clearly $\gamma$-soft, with energy minima at $\beta \sim 0.06$ and $\beta \sim 0.15$, respectively. Energy surfaces calculated with Strutinsky-BCS using deformed Woods-Saxon potential [15] also indicate $\gamma$-softness with values of $\beta_0$ comparable to the SMMC values. These calculations predict for $^{124}$Xe a prolate minimum with $\beta_0 \approx 0.20$ which is lower than the spherical configuration by 1.7 MeV but is only 0.3 MeV below the oblate saddle point, and for $^{128}$Te a shallow oblate minimum with $\beta_0 \approx 0.03$. These $\gamma$-soft surfaces are similar to those obtained in the $O(6)$ symmetry of the IBM, or more generally when the Hamiltonian has mixed $U(5)$ and $O(6)$ symmetries but a common $O(5)$ symmetry. In the Bohr Hamiltonian, an $O(5)$ symmetry occurs when the collective potential energy depends only on $\beta$ [1]. Our results are consistent with a potential energy $V(\beta)$ that has a quartic anharmonicity [2], but with a negative quadratic term that leads to a minimum at finite $\beta_0$. We have also estimated total E2 strengths from $\langle Q^2\rangle$ where $Q = e_p Q_p + e_n Q_n$ is the electric quadrupole operator with effective charges of $e_p = 1.5e$ and $e_n = 0.5e$, and extracted $B(E2; 0\to 2_1^+)$ assuming that most of the strength is in the $2_1^+$ state. We find $B(E2; 0\to 2_1^+)$ values of $663\pm 10$, $2106\pm 15$, and $5491 \pm 36$ $e^2\text{fm}^4$ to be compared with the experimental values [16] of 1164, 3910, and 9103 $e^2\text{fm}^4$ for $^{124}$Sn, $^{128}$Te and $^{124}$Xe, respectively. Thus, the SMMC calculations reproduce the correct qualitative trend; the quantitative discrepancy would be ameliorated had we used a renormalization factor similar to the potential field renormalization discussed earlier.

Information on excited states can be obtained from strength functions. The energy centroid is given by $\bar{E} = S_1/S_0$ where $S_n$ is the $n$-th moment of the relevant strength function, and can be calculated from the first logarithmic derivative of the imaginary time response function. This should be a reasonable estimate of the excitation energy of the collective state if the space of states does not include the giant resonance. We calculated the $2_1^+$ excitation energy this way from the E2 response function. The values of $1.12 \pm 0.02$, $0.96 \pm 0.02$ and $0.52 \pm 0.01$ MeV are in close agreement with the experimental values of 1.2, 0.8 and 0.6 MeV for $^{124}$Sn, $^{128}$Te and $^{124}$Xe, respectively.

Another signature of softness is the response of the nucleus to rotations. We add a cranking field $\omega J_z$ to the Hamiltonian and examine the moment of inertia as a function of the cranking frequency $\omega$. For a soft nucleus we expect a behavior intermediate between a deformed nucleus, where the inertia is independent of the cranking frequency, and the harmonic oscillator, where the inertia becomes singular. This is confirmed in Fig. 2 which shows the moment of inertia $I_2 = d\langle J_z \rangle/d\omega = \beta(\langle J_z^2 \rangle - \langle J_z \rangle^2)$ for $^{124}$Xe and $^{128}$Te as a function of $\omega$, and indicates that $^{128}$Te has a more harmonic character. The moment of inertia for $\omega = 0$ in both nuclei is significantly lower than the rigid body value ($\approx 43\hbar^2/\text{MeV}$ for $A = 124$) due to pairing correlations.

Also shown in Fig. 2 are $\langle Q^2 \rangle$ where $Q$ is the mass quadrupole, the BCS-like pairing correlation $\langle \Delta^\dagger \Delta \rangle$ for the protons and $\langle J_z \rangle$. Notice that the increase in $I_2$ as a function of $\omega$ is strongly correlated with the rapid decrease of pairing correlations and that the peaks in $I_2$ are associated with the onset of a decrease in collectivity (as measured by $\langle Q^2 \rangle$). This suggests band crossing along the Yrast line associated with pair breaking and alignment of the quasi-particle spins at $\omega \approx 0.2$ MeV ($\langle J_z \rangle \approx 7\hbar$) for $^{128}$Te and $\omega \approx 0.3$ MeV ($\langle J_z \rangle \approx 11\hbar$) for $^{124}$Xe. Our results are consistent with an experimental evidence of band crossing in the Yrast sequence of $^{124}$Xe around spin of 10 $\hbar$ [17]. The alignment effect is clearly seen in the behavior of $\langle J_z \rangle$ at the lower temperature which shows a rapid increase after an initial moderate change. Deformation and pairing decrease also as a function of temperature.

We have analyzed the number of correlated pairs of these nuclei in their ground state. For a given angular momentum $J$, we define the pair operators $A_{JM}^\dagger(ab) = 1/\sqrt{1+\delta_{ab}}[a_{j_a}^\dagger \times a_{j_b}^\dagger]^{JM}$. These operators are boson-like in the sense that they satisfy the expected commutation relations in the limit when the number of valence nucleons is small compared with the total number of single-particle states in the shell. In the SMMC we compute the pair correlation matrix in the ground state $\sum_M \langle A_{JM}^\dagger(ab) A_{JM}(cd) \rangle$, which is an hermitian and positive-definite matrix in the space of ordered orbital pairs $(ab)$ (with $a \leq b$). This matrix can be diagonalized to find the eigenbosons $B_{\alpha JM}^\dagger = \sum_{ab} \psi_{\alpha J}(ab) A_{JM}^\dagger(ab)$, where $\alpha$ labels the various bosons with the same angular momentum $J$. These eigenbosons approximately satisfy

$$\sum_M \langle B_{JM}^\dagger(\alpha) B_{JM}(\gamma) \rangle = n_J(\alpha) \delta_{\alpha\gamma} , \qquad (7)$$

where the positive eigenvalues $n_{\alpha J}$ are the number of $J$-pairs of type $\alpha$. We have calculated the total number of $J$-pairs ($n_J = \sum_\alpha n_{\alpha J}$) in the various pairing channels, and the results for the number of correlated pairs (after subtracting out the mean-field values [18]) are shown in Fig. 3. Since the number of neutrons in $^{124}$Xe is above 66, they are treated as holes. For $J = 0$ and $J = 2$ we can compare the largest $n_{\alpha J}$ with the number of $s$ and $d$ bosons obtained from the $O(6)$ limit of the IBM. In the latter we use the exact $O(6)$ formula [3] for the average number of pairs and multiply by the relative fraction of protons and neutrons to find the pair content for each type of nucleon. For $^{124}$Xe the SMMC (IBM) results in the proton-proton pairing channel are 0.85 (1.22) $s$ ($J = 0$) pairs, and 0.76 (0.78) $d$ ($J = 2$) pairs, while in the neutron-neutron channel we find 1.76 (3.67) $s$ pairs and 2.14 (2.33) $d$ pairs. For the protons the SMMC $d$ to $s$ pair ratio 0.89 is close to its $O(6)$ value of 0.64. However, the same ratio for the neutrons, 1.21, is intermediate between $O(6)$ and $SU(3)$ (where its value is 1.64) and is consistent with the neutrons filling the middle of the shell. The total numbers of $s$ and $d$ pairs – 1.61 proton pairs and 3.8 neutron (hole) pairs – are below the IBM values of 2 and 6, respectively. The fermion-based SMMC calculations indeed indicate pair correlations for higher $J$ values.

In conclusion, we have presented the first microscopic evidence of softness in nuclei in the 100-140 mass region, using the SMMC for the full 50-82 major shell. Work in progress on the systematic in this region would include a $Q_p \cdot Q_n$ force [5]. However, such calculations are more time-consuming as this interaction violates the Monte Carlo sign rule.

This work was supported in part by the Department of Energy, Contracts No. DE-FG-0291-ER-40608 and DE-FG0690-ER40561, and by the National Science Foundation, Grants No. PHY90-13248 and PHY91-15574. We acknowledge useful discussions with F. Iachello, H. Nakada and A. Ansari. YA acknowledges the hospitality of the INT at Seattle where part of this work was completed. Research sponsored in part by the Phillips Laboratory, Air Force Materiel Command, USAF, under cooperative agreement number F29601-93-2-0001.

FIG. 1. Free energy surfaces for $^{128}$Te (left) and $^{124}$Xe (right) in the $\beta - \gamma$ plane at several temperatures. The contour lines are separated by 0.3 MeV and the lighter shades correspond to lower energies. Notice the $\gamma$-softness of the surfaces.

FIG. 2. Observables for $^{124}$Xe and $^{128}$Te as a function of cranking frequency $\omega$ and for two temperatures. $I_2$ is the moment of inertia, $Q$ is the mass quadrupole moment, $\Delta$ is the $J = 0$ pairing operator, and $J_z$ is the angular momentum along the cranking axis.

FIG. 3. Number of correlated pairs of angular momentum $J$ in $^{124}$Xe. Different shades correspond to p-p, n-n, isoscalar and isovector p-n pairs. Neutrons are treated as holes.